\documentclass[%
]{article}

\usepackage{amssymb}
\usepackage{amsmath,amsfonts,amsthm,bm}
\usepackage{upgreek}
\usepackage{mathrsfs}
\usepackage{caption}
\usepackage{epstopdf}
\usepackage{tabularx}
\usepackage{array}
\usepackage{siunitx}
\usepackage{subcaption}
\usepackage{mathtools}
\usepackage{amsthm}
\usepackage[round]{natbib}
\usepackage{amsmath,etoolbox}
\usepackage{xcolor}
\usepackage{soul}
\usepackage{enumerate}
\usepackage{authblk}
\usepackage[
    letterpaper,
    top=2cm,
    bottom=2cm,
    left=2cm,
    right=2cm,
    textwidth=16cm,
    textheight=23cm
]{geometry}
\usepackage{hyperref}
\hypersetup{
    colorlinks=true,
    linkcolor=blue,
    citecolor=black,
    filecolor=magenta,      
    urlcolor=cyan,
}

\theoremstyle{plain}
\newtheorem{theorem}{Theorem}
\theoremstyle{definition}

\newtheorem{remark}[theorem]{Remark}

\newcommand{\jump}[1]{\left[\kern-0.15em\left[ #1 \right]\kern-0.15em\right]}  
\newcommand{\norm}[1]{\vert #1 \vert}

\newcommand{\I}{\mathbf{I}}

\newcommand{\nv}{\mathbf{n}}

\newcommand{\Nv}{\boldsymbol{\mathcal{N}}}

\newcommand{\Xv}{\mathbf{X}}
\newcommand{\xv}{\mathbf{x}}
\newcommand{\yv}{\mathbf{y}}

\newcommand{\Fb}{\mathbf{F}}
\newcommand{\Rb}{\mathbf{R}}
\newcommand{\Ub}{\mathbf{U}}

\newcommand{\Cb}{\bm{\mathbb{C}}}

\newcommand{\uvb}{\mathbf{u}}

\newcommand{\bvb}{\mathbf{b}}
\newcommand{\hvb}{\mathbf{h}}
\newcommand{\mvb}{\mathbf{m}}
\newcommand{\Bvb}{\mathbf{B}}

\newcommand{\Hvb}{\mathbf{H}}
\newcommand{\Svb}{\mathbf{S}}
\newcommand{\Mvb}{\mathbf{M}}
\newcommand{\sigvb}{\boldsymbol{\sigma}}

\newcommand{\Tmech}{\mathbf{T}}

\newcommand{\hvbr}{\mathbf{h}^r}
\newcommand{\Bvbr}{\boldsymbol{\mathcal{B}}^r}
\newcommand{\Hvbr}{\boldsymbol{\mathcal{H}}^r}
\newcommand{\Hvbrdot}{\dot{\boldsymbol{\mathcal{H}}}^r}

\newcommand{\muo}{\mu_{\texttt{0}}}

\newcommand{\Rthr}{\mathbb{R}^3}

\newcommand{\mt}{\texttt{m}}
\newcommand{\pt}{\texttt{p}}

\newcommand{\mecht}{\texttt{mech}}
\newcommand{\magt}{\texttt{mag}}

\newcommand{\cplt}{\texttt{couple}}

\newcommand{\p}{\partial}

\newcommand{\Vo}{{\mathcal{V}_0}}
\newcommand{\V}{{\mathcal{V}}}

\newcommand{\dV}{{\partial \mathcal{V}}}

\definecolor{mypink}{RGB}{180, 48, 122}

\begin{document}

\title{Stretch-independent magnetization in \\ incompressible magnetorheological elastomers}

\author[1,2]{Kostas Danas\thanks{Email: \texttt{konstantinos.danas@polytechnique.edu}}}
\author[3]{Pedro M. Reis\thanks{Email: \texttt{pedro.reis@epfl.ch}}}

\affil[1]{\textit{LMS, CNRS, \'Ecole Polytechnique, Institut Polytechnique de Paris, Palaiseau, 91128, France}}
\affil[2]{\textit{ELyTMaX, CNRS, Tohoku University, Sendai, Japan}}
\affil[3]{\textit{Flexible Structures Laboratory, Institute of Mechanical Engineering, \'Ecole Polytechnique F\'ed\'erale de Lausanne (EPFL), 1015 Lausanne, Switzerland }}





\maketitle 

\begin{abstract}
In this study, we perform a critical examination of the phenomenon where the magnetization is stretch independent in incompressible hard-magnetic magnetorheological elastomers ($h$-MREs), as observed in several recent experimental and numerical investigations. We demonstrate that the fully dissipative model proposed by \cite{mukherjee2021} may be reduced, under physically consistent assumptions, to that of \cite{Yan2023}, but not that of \cite{zhao2019}. In cases where the $h$-MRE solid undergoes non-negligible stretching, the model of \cite{zhao2019} provides predictions that are in disagreement with experimental observations given that, by construction, that model produces a magnetization response that is not stretch-independent. By contrast, the other two models are able to describe this important feature present in $h$-MREs, as well as in incompressible magnetically soft $s$-MRES. Note that in cases where stretching is negligible, such as for inextensible slender structures under bending deformation, the \cite{zhao2019} model provides accurate predictions despite its underlying assumptions. Additionally, our analysis reveals two key points about the magnetization vector in the context of the more general, fully dissipative model. First, the magnetization can be \textit{related} to an internal variable in that theory. However, it cannot be formally used as an internal variable except in the special case of an ideal magnet, and, as such, it is subject to constitutive assumptions. Furthermore, we clarify that the magnetization vector alone is insufficient to describe entirely the magnetic response of an MRE solid; instead, the introduction of one of the original Maxwell fields is always necessary for a complete representation.\\


\end{abstract}

\section{Introduction and problem definition}
\label{sec:Introduction and problem definition}

In light of the recently burgeoning interest in magneto-elastic materials, a plethora of
theoretical, numerical, and experimental studies have emerged in the literature on magnetically soft ($s$-) and hard ($h$-) magnetorheological elastomers (MREs), also known as magnetoactive elastomers or polymers.
In the laboratory, three major classes of MREs have been fabricated: $s$-MREs containing carbonyl-iron particles (or other low dissipative ferromagnetic particles), $h$-MREs comprising magnetically dissipative particles (such as NdFeB or similar), and hybrid MREs combining both types of particles. In most cases, the micron-sized particles are in the form of a powder and have fairly spherical or polyhedral shapes \citep{schmann2017,schumann2017characterisation}. 

The $s$-MREs exhibit high magnetic permeability and magnetization saturation but demagnetize immediately after removal of the magnetic field \citep{danas12}. By contrast, $h$-MREs have a significant magnetic coercivity and smaller magnetic saturation but retain their magnetization upon removal of the magnetic field \cite{Stepanov2017}. At this pre-magnetized state, however, $h$-MREs usually exhibit a relatively low magnetic permeability, close to unity. Soft- and hard-magnetic particles can be combined in a matrix, yielding a new class of hybrid MREs \citep{Stepanov2017,Moreno2022npj},  that combine, in a non-trivial manner, the individual advantages of $s$-MREs and $h$-MREs and improve the coupled magneto-mechanical response. 
With the exception of $s$-MREs, which can be cured under a magnetic field to form particle chains and thus exhibit mechanical and magnetic anisotropy \citep{danas12}, the majority of the magneto-elastic material systems fabricated to date tend to be mostly isotropic, appearing nearly homogeneous at the macroscopic scale. 
Incompressible MREs exhibit maximum 
magnetostriction, and thus, most past studies have focused on the incompressible limit.

Here, we discuss a critical observation relevant to incompressible $s$- and $h$-MREs, focusing on the \emph{experimentally observed} stretch-independence of the magnetic response of these materials. During the past decade, several studies, theoretical \citep{mukherjee2021} and experimental \cite{danas12} for $s$-MREs and \cite{Yan2021ijss,Yan2023} for $h$-MREs, have demonstrated that the amplitude of the magnetization is independent of the stretching (or stressing) of the material. This finding distinctly opposes the well-known magneto-elastic Villari effect observed in the context of pure metallic polycrystalline magnets \citep{Kuruzar1971,daniel2014}, where the application of stress leads to a change of the magnetization response, both the magnetic permeability and magnetization saturation of the magnet. By contrast, the stretch independence of MREs has important consequences for their response when actuated by external magnetic fields. 

In the past few years, numerous studies on $h$-MREs have made extensive usage of the \cite{zhao2019} model, which has been mainly tested against experimental data on slender structures subjected to \textit{pure} bending and in the absence of any mechanical pre-stresses or pre-stretches (at least of a non-negligible amplitude). In that model and several studies thereafter, the authors make the assumption that the initial pre-magnetization, or, more precisely, remanent magnetic flux, transforms with the deformation gradient. This assumption directly implies that the magnetization of the $h$-MRE will change upon application of tensile or compressive loads, which, as we shall review below, is not supported by recent \citep{Yan2023} and earlier \citep{danas12} experimental and numerical studies \citep{mukherjee2021}. By contrast, the models of \cite{mukherjee2021,mukherjee2022} and \cite{Yan2021ijss,Yan2023} for $h$-MREs and the former models of \cite{danas12}, \cite{Lefevre2017} and \cite{mukherjee2020} for $s$-MREs account for this stretch independence of the magnetization response to a fair extent and, thus, are able to describe cases with non-zero pre-stresses and pre-stretches predictively. 

The main focus of the present study is to closely examine and clarify the similarities and differences between (i) the simpler uncoupled models of \cite{Yan2023} and \cite{zhao2019} for $h$-MREs and small magnetic loads around the pre-magnetization state and (ii) the fully dissipative coupled model of \cite{mukherjee2021}. The latter is pertinent for both $s$-and $h$-MREs as well as magneto-mechanical loads of arbitrary amplitude. By coupled and uncoupled magneto-mechanical response, we refer to the ability of the model to predict intrinsic magnetostriction of an MRE under Eulerian applied magnetic fields in the sense described by \cite{danas2017effective} and clearly discussed in Section~\ref{sec:An energetic model for small magnetic fields after pre-magnetization}. More importantly, we will show that under certain, physically sound assumptions, the \cite{mukherjee2021} model may be reduced to the \cite{Yan2023} model but not that of \cite{zhao2019}.

Our manuscript is organized as follows. In Section~\ref{sec:Preliminary definitions}, we introduce the main mechanical and magnetic quantities needed for the analysis of the magneto-mechanical problem. In Section~\ref{sec:The magnetic dissipation model of Mukherjee et al.}, we recall, briefly and concisely, the main ingredients of the fully dissipative model of \cite{mukherjee2022} in the $\Fb$-$\Bvb$ space, which is an exact Legendre dual of the original model of \cite{mukherjee2021} that was proposed in the $\Fb$-$\Hvb$ space. Here, $\Fb$ denotes the deformation gradient, while $\Bvb$ and $\Hvb$ are the Lagrangian magnetic flux and field strength, respectively. In Section~\ref{sec:An energetic model for small magnetic fields after pre-magnetization}, we simplify, under physically sound assumptions, the previous fully dissipative model to two simpler coupled and uncoupled energetic models for $h$-MREs, both of which are valid for small applied magnetic fields around the pre-magnetized state. Then, we summarize the $h$-MRE model of \cite{Yan2023}, showing that it turns out to be identical to the uncoupled energetic model of the present note. We proceed by discussing connections and differences between the former two rotation-based models and the \cite{zhao2019} model. Finally, we close by discussing the limitations of the simpler models as well as the effect that the modeling of the surrounding air has on the response of the MRE.

\section{Preliminary definitions}
\label{sec:Preliminary definitions}

We consider a magnetoelastic deformable solid that occupies a region $\Vo$ (or $\V$) with boundary $\p\Vo$ (or $\p\V$) of outward normal $\Nv$ (or $\nv$) in the undeformed stress-free (or current) configuration. Material points in the solid are identified by their initial position vector $\Xv$ in the undeformed configuration $\Vo$, while the current position vector of the same point in the deformed configuration $\V$ is given by $\xv=\yv(\Xv)=\Xv+\uvb(\Xv)$, with $\uvb$ denoting the displacement vector. Motivated by the usual physical arguments, the mapping $\yv$ is required to be continuous and one-to-one on $\Vo$. In addition, we assume that $\yv$ is twice continuously differentiable, except, possibly, on existing interfaces (\textit{e.g.}, due to the presence of different phases) inside the material. The deformation gradient is then denoted by $\Fb=\text{Grad} \yv=\I+\text{Grad}\uvb$ and its determinant $J=\det \Fb >0$ with $\I$ being the second-order identity tensor. Moreover, ``Grad'' denotes the gradient operator with respect to $\Xv$ in the reference configuration. In addition, the reference density of the solid $\rho_0$ is related to the current density $\rho$ by $\rho_0=\rho J$. Time dependence in not considered here. 

Traditionally, in the absence of electric currents and charges, the following three quantities are used to describe the magnetic state of a solid in the \emph{current} configuration:
\begin{itemize}
\item the current magnetic flux $\bvb$;
\item the current magnetic field strength $\hvb$; and
\item the current magnetization $\mvb$, which, by construction, is zero in non-magnetic domains.
\end{itemize}
It is important to note, however, that these three quantities are not independent of one another; they are related by the \emph{constitutive} relation $\bvb=\mu_0 (\hvb+\mvb)$, which may be recast as 
\begin{equation}
	 \mvb=\dfrac{1}{\mu_0}\bvb-\hvb \qquad \mathrm{in} \quad \V, 
	\label{eq:magnetization_definition}
\end{equation}
with $\mu_0$ denoting the magnetic permeability of vacuum, air or non-magnetic solids. 

\begin{remark}
In fact, the expression in Eq.~\eqref{eq:magnetization_definition} is a \emph{definition} of the magnetization vector in the \textit{current} volume $\V$, which, however, is not defined on its boundary $\dV$, and does not have a unique Lagrangian definition \citep{dorfmann2004nonlinear}. Moreover, by definition, $\mvb=\mathbf{0}$ in a non-magnetic body. This statement implies that $\mvb$ is insufficient as a variable to describe the presence of magnetic lines (in the sense of Maxwell) in the surrounding air or in the non-magnetic medium more generally (\textit{e.g.,} a polymer), two settings that are commonly of interest in most problems involving magnetic materials. In these two cases, one is left with the relation $\bvb=\mu_0\hvb$, which implies (in the sense of a continuum medium) that $\bvb$ is linearly dependent on $\hvb$, and \emph{vice versa}, via the magnetic constitutive parameter $\mu_0$. Henceforth, we seek to reconcile, in certain special cases, the approaches using the original Maxwell fields $\bvb$ and $\hvb$ (or their Lagrangian counterparts discussed below) as working variables \citep{dorfmann2003magnetoelastic} and those using all three, $\bvb$, $\hvb$, and $\mvb$ \citep{brown1963micromagnetics,JamesKinder93,kankanala2004finitely}. Moreover, in Section~\ref{sec:An energetic model for small magnetic fields after pre-magnetization}, we will show that $\mvb$, unlike the original Maxwell fields $\bvb$ and $\hvb$, can be related to an internal--and not an independent--variable that describes permanent magnetization states in the MRE solid.  
\end{remark}

At large strains, the fields $\bvb$ and $\hvb$ can be pulled back from $\V$ to $\Vo$ to their Lagrangian forms, denoted by $\Bvb$ and $\Hvb$, respectively, such that \citep{dorfmann2003magnetoelastic,bustamante2008}
\begin{equation}
	\Bvb=J \Fb^{-1} \bvb, \qquad {\rm and} \qquad \Hvb=\Fb^T \hvb.
	\label{eq:transforms_B_and_H}
\end{equation}
Moreover, as has been extensively discussed in the literature (see, for instance, \cite{dorfmann2005some}), Eq.~\eqref{eq:magnetization_definition} is not form invariant under transformations, which is a manifestation of the non-unique definition of $\mvb$. The Lagrangian $\Bvb$ is also divergence-free and $\Hvb$ is curl-free, such that
\begin{equation}
	 \mathrm{Div}\Bvb=0 \quad \mathrm{in}\ \Vo, \qquad \quad \jump{\Bvb}\cdot\Nv=0 \quad \mathrm{in} \ \dV_0^\mathtt{B},
	 \label{eq:divb_bdotn_balance_law_reference_def}
\end{equation}
and
\begin{equation}
	\mathrm{Curl}\,\Hvb = \mathbf{0}\quad \mathrm{in}\ \Vo, \qquad \quad \jump{\Hvb}\times \Nv = \mathbf{0} \quad \mathrm{on}\ \dV_0^\mathtt{H}.
	\label{eq:Ampere_law_pde_reference}
\end{equation}
The explicit notation $\dV_0^\mathtt{B}$ and $\dV_0^\mathtt{H}$ serve to denote the corresponding parts of the boundary where jumps in $\Bvb$ and $\Hvb$ are applied.

The total Cauchy stress tensor $\sigvb$ and the total (first) Piola-Kirchhoff $\Svb$ read, respectively,
\begin{equation}
\sigvb = \dfrac{1}{J} \Svb \Fb^{T}, \quad \mathrm{ and} \quad \Svb=J \Fb^{-1}\sigvb. \label{eq:transforms_stress}
\end{equation} 
Both of these stress measures are divergence free, for instance,
\begin{equation}
\mathrm{Div} \ \Svb = \bm{0} \quad \text{in} \ \Vo,  \quad \Svb \Fb^{T} = \Fb \Svb^T, \quad \text{and} \quad \jump{\Svb}\cdot \bm{\mathcal{N}}-\Tmech = \bm{0} \quad \text{on} \ \dV_0^\mathtt{N}, 
\end{equation}
where $\Tmech$ denotes the mechanical traction in the reference configuration applied on the corresponding part of the boundary $\dV_0^\mathtt{T}$. 

\section{The magnetic dissipation model of \cite{mukherjee2021}}
\label{sec:The magnetic dissipation model of Mukherjee et al.}

\subsection{Internal variable for magnetic dissipation}
\label{sec:Internal variable for magnetic dissipation}

A thermodynamically consistent model for any dissipative material may be constructed through the definition of a finite number of internal variables, which reflect the irreversible processes the material undergoes under external loads. Those internal variables are, in general, difficult to measure directly in an experiment (\textit{e.g.}, plastic strain, or magnetization), and moreover, they cannot be controlled with direct manipulations \citep{bassiouny1988, eringenmaugin90}. Nevertheless, they are necessary to describe the time evolution of the internal state of the material since they carry information on the history of the processes (\textit{e.g.}, motion of dislocations and domain walls, bypassing or pinning at obstacles). In this regard, one of the main differences between $h$-MREs and $s$-MREs is the underlying magnetic dissipation of the filler particles (\textit{e.g.}, NdFeB) in the former. Upon cyclic magnetic loading, as a consequence of the finite strains and the magneto-mechanical coupling, the response of the $h$-MRE composite exhibits both magnetic and mechanical hysteresis.

\begin{figure}[h!]
    \centering
    \includegraphics[width=0.75\columnwidth]{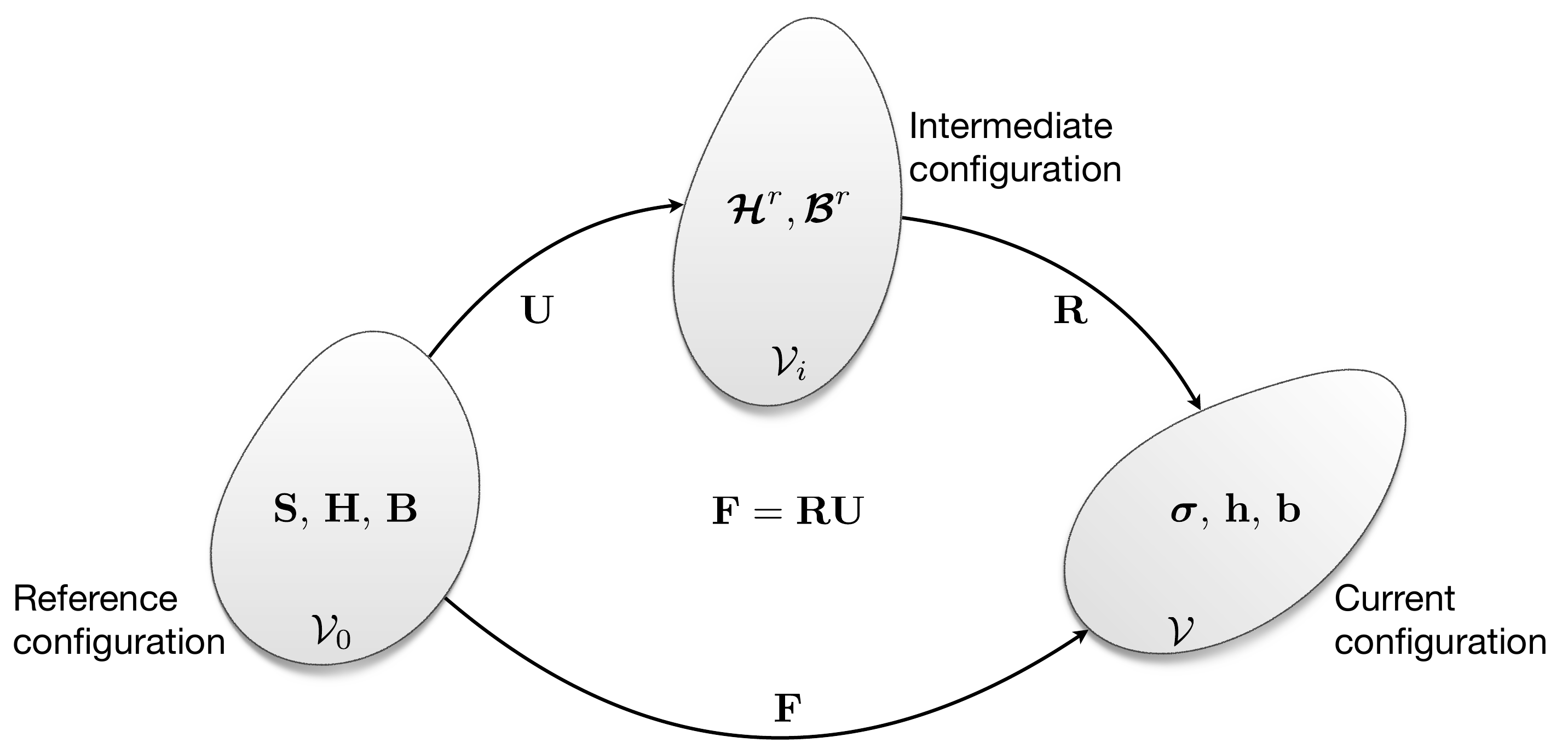}
    \caption{Definition of the reference, intermediate and current configurations of volume $\Vo$, $\mathcal{V}_i$ and $\V$, respectively, along with the different field variables defined therein.}
    \label{fig:configurationsFH}
\end{figure}

Through the extensive analysis of data from numerical RVE simulations,
\cite{mukherjee2021} have shown that only one internal remanent $H$-like\footnote{Given the arbitrary nature of choosing internal variables, it is evident that one could also have reasonably chosen a $B$-like variable. \cite{mukherjee2022} subsequently showed that such a choice is inconsequential for the analysis and is a mere matter of \textit{taste}.} vector variable,
\begin{align}
	\Hvbr\in\mathbb{R}^3,
	\label{eq:Hvbr_def}
\end{align}
which lies in the \emph{stretch-free}, \emph{intermediate} configuration $\mathcal{V}_i$ (see Fig.~\ref{fig:configurationsFH}) suffices to describe the magneto-mechanical behavior of an \emph{incompressible} $h$-MRE. This assumption is not unusual. For example, the analogous classical $J_2$ flow theory of elasto-plasticity makes exactly the same assumptions for the plastic deformation gradient $\Fb^p$ \citep{Lee1969} or plastic strain $\boldsymbol{\varepsilon}^p$ \citep{Hill1950}. 

A direct consequence of the above unambiguous numerical and theoretical results is that the current magnetization vector, $\mvb$ is affected by macroscopic rotations but not stretches, as we will show in the following sections. These observations are also confirmed experimentally and independently in the works of \cite{Yan2021ijss} and \cite{Yan2023} for $h$-MREs and the earlier study of \cite{danas12} for $s$-MREs\footnote{We recall that the case of $s$-MREs may be recovered in the limit of vanishing dissipation of the $h$-MRE model \citep{Danas2024}.}. 
In particular, \cite{Yan2023} have first employed a three-dimensional (3D) model, based on that of \cite{zhao2019} (hereon denoted as the $\Fb\Mvb$-model with $\Mvb$ being the pre-magnetization vector), which leads to changes of the magnetization amplitude (and direction) upon application of a deformation gradient. The authors showed that, in some cases where stretching is not negligible, this $\Fb\Mvb$ model yields unphysical predictions (\textit{i.e.}, in disagreement with experimental data), as they demonstrated for the specific configuration of a thin plate made of an $h$-MRE, actuated under combined mechanical (pressure) and magnetic loading. 
Following an alternative point of departure, these authors then developed a model (hereon denoted as the $\Rb\Mvb$-model), which, upon application of a deformation gradient, $\Fb=\Rb\Ub$, imposes changes \emph{solely} in the orientation of the magnetization vector due to the presence of only the rotation part, $\Rb$, and \textit{not} the stretch part, $\Ub$. This $\Rb\Mvb$-model was found to yield predictions in excellent agreement with experimental data, as we will discuss in more detail in Section~\ref{sec:Connections and differences between the models}. \cite{Yan2023} also demonstrated that classical dimensional-reduction techniques to derive equilibrium equations for \textit{inextensible} slender structures (beams, elastica, plates, and shells) that invoke Euler-Bernoulli or Kirchhoff-Love hypotheses for the kinematics (\textit{i.e.}, normals stay normal to the center-line/mid-surface and do not stretch) naturally \textit{block} the stretching and give the erroneous impression that $\Fb\Mvb$ models are appropriate for such structures made of $h$-MREs. Similarly, but for incompressible $s$-MREs, earlier experimental work by \cite{danas12} showed that pre-stressing of the material (even when anisotropic) and upon the application of an external magnetic field, leads to an effectively unchanged amplitude of the magnetization response, albeit strongly affecting its magnetostriction.

The two independent experiments discussed above and reported in \cite{Yan2023} and \cite{danas12}, together with their respective numerical and theoretical analyses therein, provide unambiguous and convincing evidence that the \emph{amplitude of the magnetization} response of \emph{incompressible} MREs more generally is insensitive to mechanical stretches. 
Certainly, considering large shear loads may alter the direction of magnetization, consequently rendering the analysis highly nuanced and intricate.  In this context, it is noteworthy that incompressible MREs do not exhibit the inverse magnetoelastic Villari effect of metallic magnets, which do so but are compressible \citep{Kuruzar1971}.
\begin{remark}
Both push-forward or pull-backward transformations of the internal variable $\Hvbr$ may always be considered, resulting in two additional measures: one Eulerian, $\hvbr=\Rb\Hvbr$, and the other Lagrangian, $\Hvb^r = \Ub \Hvbr$, respectively. Nevertheless, those measures are non-essential since, given that they are not independent variables, they cannot be given any particular physical interpretation, and more importantly, they cannot be measured or directly controlled \citep{eringenmaugin90}. 
\end{remark}
 
In Section~\ref{sec:An energetic model for small magnetic fields after pre-magnetization}, we will demonstrate that the current magnetization $\mvb$ is a function of $\Hvbr$, implying that $\mvb$ (or its representation in a different configuration) can be linked directly to an internal variable; cf. the relevant work of \cite{klinkel2006phenomenological, linnemann2009constitutive,kalina2017modeling}). However, we will see that failing to use an appropriate transformation of $\mvb$ can lead to important errors. We will then establish a direct connection between the full dissipative theory of \cite{mukherjee2021} and \cite{mukherjee2022}, as well as the simplified energetic models of \cite{Yan2021ijss} and \cite{Yan2023}. We emphasize that the latter two studies provided strong experimental evidence for the fact that the current magnetization is stretch-independent and affected only by rotations.

\subsection{The isotropic magneto-mechanical invariants for $h$-MREs}
\label{sec:The isotropic magneto-mechanical invariants for hMREs}

A natural way to satisfy the conditions of even magneto-mechanical coupling, isotropic material symmetry, and frame indifference is to express the energy density and dissipation in terms of appropriately chosen \emph{isotropic} invariants. In \cite{mukherjee2022}, a dual formulation in the sense of Legendre-Fenchel was proposed for $h$-MREs, yielding exactly equivalent constitutive laws in both the $\Fb$-$\Hvb$ and $\Fb$-$\Bvb$ space. In the present study, we focus on the $\Fb$-$\Bvb$ formulation, which allows us to make direct contact with \cite{Yan2023}, as well as assess the limitations of the earlier model of \cite{zhao2019}. Moreover, for consistency with earlier works by a subset of the authors of these two studies, we will also present a quasi-incompressible version of the models. Note, however, that the formalism we will propose is only valid for minor volume changes and not for general compressible MREs; for the latter, we refer to the recent work of \cite{Gebhart2022a,Gebhart2022b} on this topic.

First, we will define the general set of available invariants, $\Cb=\Fb^T\Fb$, $\Bvb$, and $\Hvbr$, given the corresponding arguments. Subsequently, we will select a subset of them to model the $h$-MREs; a choice that is primarily motivated by corresponding numerical RVE simulations of two-phase $h$-MRE composites 
\citep{mukherjee2021}. While this choice does not represent a rigorous result, it serves as an effective homogenization-guided approach. This strategy ensures that the number of invariants remains minimal and keeps the model entirely explicit. 

\vspace{0.2in}
{\emph{Mechanical invariants}.} 
\begin{align}
\label{eq:mech_invars}
    I_1 = \text{tr}(\Cb), \qquad I_3 = J^2 = \det \Cb=1,
\end{align}

{\emph{Magneto-mechanical invariants in $\Fb$-$\Bvb$ formulation}.} 
\begin{align}
\label{eq:FB_invariants}
&I_4^{\mathtt{B}} = \Bvb \cdot \Bvb, & 
&I_4^{\mathtt{BHr}} = \Bvb \cdot \Cb^{-1/2} \Hvbr, & 
&I_4^{\mathtt{Hr}} = \Hvbr \cdot \Cb \Hvbr \nonumber \\
&I_5^{\mathtt{B}} = \Bvb \cdot \Cb \Bvb , & 
&I_5^{\mathtt{BHr}} = \Bvb \cdot \Cb^{1/2}\Hvbr, & 
&I_5^{\mathtt{Hr}} = \Hvbr \cdot \Hvbr. \nonumber\\
&I_6^{\mathtt{B}} = \Bvb \cdot \Cb^2 \Bvb, &
&I_6^{\mathtt{BHr}} = \Bvb \cdot \Cb^{3/2}\Hvbr  & 
&I_6^{\mathtt{Hr}} = \Hvbr \cdot \Cb^2 \Hvbr. 
\end{align}

\subsection{Energy densities and dissipation potential}
\label{subsec:Energy densities and dissipation potential}

We express the energy density, $W(\Fb,\Bvb)$, as the sum of three distinct energy densities, namely, the purely mechanical ($\rho_0\Psi_\mecht$), purely magnetic ($\rho_0\Psi_\magt$) and coupling ($\rho_0\Psi_\cplt$), such that\footnote{In the original work of \cite{mukherjee2022} the notation $W^{\mathtt{B}}$ was used to distinguish between the $\Fb$-$\Bvb$ model and the equivalent dual in the $\Fb$-$\Hvb$ space denoted with $W^{\mathtt{H}}$. In the present work, we only use the $\Fb$-$\Bvb$ version, and thus, the relevant superscripts will be dropped for simplicity of the notation. In turn, for clarify, the superscripts will be maintained when writing the invariants.}
\begin{align}
W(\Fb,\Bvb,\Hvbr) = \rho_0 \Psi_{\mathtt{mech}}(I_1,J) + \rho_0 \Psi_{\magt}(I_5^{\mathtt{B}},I_5^{\mathtt{BHr}}, I_5^{\mathtt{Hr}})+\rho_0 \Psi_{\cplt} (I_4^{\mathtt{Hr}}, I_5^{\mathtt{BHr}}, I_5^{\mathtt{Hr}},I_6^{\mathtt{BHr}}) + \dfrac{1}{2 \mu_0 J}  I_5^{\mathtt{B}}, \label{eq:WB_full}
\end{align}
where $\rho_0$ is the reference density of the solid, and the last term ($I_5^{\mathtt{B}}/2\mu_0$) in Eq.~\eqref{eq:WB_full} represents the energy associated with free space with $\mu_0$ being the magnetic permeability in vacuum or in non-magnetic solids such as the polymer matrix phase. This last term is necessary for mathematical consistency \citep{dorfmann2003magnetoelastic} as well as for modeling the effect of the surrounding air upon the MRE body. We also note that in the proposed model in Eq.~\eqref{eq:WB_full}, a subset of the invariants defined in Eq.~\eqref{eq:FB_invariants} was found to be sufficient for the problem at hand.   

\paragraph{The mechanical energy density}
The purely mechanical free energy density $\rho_0 \Psi_{\mecht}$ in Eq.~\eqref{eq:WB_full} may be chosen to correspond to the analytical homogenization estimate of \citet{lopezpamies2013} for a two-phase composite made of an incompressible nonlinear elastic matrix comprising isotropic distributions of rigid-particles, such that
\begin{equation}
\rho_0 \Psi_{\mecht}(I_1,J) = (1-c) \rho_0 \Psi_{\mathtt{m,mech}}(\mathcal{I}_1) + \dfrac{K_{\mathtt{m}}}{2 (1-c)^6}(J-1)^2, \qquad \mathcal{I}_1 = \dfrac{J^{-2/3} I_1-3}{(1-c)^{7/2}}+3, \label{eq:Psi_mech}
\end{equation}
where $c$ is the particle volume fraction, $K_{\mathtt{m}}$ (much larger than the shear modulus) is the compressibility modulus  and $\Psi_{\mathtt{m,mech}}$ is the free energy density of the matrix.
The purely incompressible result (\textit{i.e.},  $K_\mt \to \infty$) recovers the dilute estimate of \cite{Einstein1906} and satisfies the well-known \cite{Hashin1963} bounds for such composites. Notably, the homogenization estimate in Eq.~\eqref{eq:Psi_mech} holds for any $I_1$-based incompressible rigid-particle--matrix composite and was shown by \cite{Luo2023} to also be extremely accurate for quasi-incompressible matrices. Thus, the choice for the constitutive law of the matrix phase remains versatile in the present modeling framework. Evidently, in the limit of $c=0$, the homogenized energy recovers that of the matrix phase; \textit{i.e.}, $\lim_{c\to 0}\Psi_{\mecht}(I_1,J)=\Psi_{\mathtt{m,mech}}(I_1,J)$. By contrast, $\lim_{c\to 1}\Psi_{\mecht}(I_1,J) = +\infty$, thus recovering the energy of a mechanically rigid material, such as that of the particle in the present case. 

It is important to note that the above homogenized mechanical energy for the MRE may be replaced readily by any other mechanical energy of a phenomenological type that is available or better suited for the material at hand, as will be discussed in Section~\ref{sec:The RM energetic model of Yan et al.}.

\paragraph{The magnetic energy}
The magnetic free energy, $\rho_0\Psi_{\magt}$, in Eq.~\eqref{eq:WB_full} reads \citep{mukherjee2022}
\begin{align}
\rho_0 \Psi_{\magt}(I_5^{\mathtt{B}},I_5^{\mathtt{BHr}}, I_5^{\mathtt{Hr}}) = - \dfrac{1}{2\mu_0} \dfrac{ \chi^{e} }{1+\chi^{e}} I_5^{\mathtt{B}} +  I_5^{\mathtt{BHr}} + \dfrac{\mu_0}{2} \bigg( \chi^e + \dfrac{1+2c}{3c} \bigg) I_5^{\mathtt{Hr}}+ \dfrac{\mu_0}{c} \dfrac{(m^s)^2}{\chi^r_{\mathtt{p}}} f_{\mathtt{p}} \bigg( \dfrac{\sqrt{I_5^{\mathtt{Hr}}}}{m^s} \bigg).  \label{eq:magnetic_energy_FB}
\end{align}
In Eq.~\eqref{eq:magnetic_energy_FB}, $\chi^r_\pt$ is the \emph{remanent susceptibility} of the underlying magnetic particles, whereas the ``effective'' parameters $\chi^e$ and $m^s$ for the composite are given in terms of the particle magnetic properties and its volume fraction $c$ as 
\begin{equation}
\chi^e = \dfrac{3c \chi^e_\pt}{3+(1-c) \chi^e_\pt}, \qquad \text{and }\qquad  m^s = c \ m^s_\pt \bigg( \dfrac{1+\chi^e_\pt }{1 + \chi^e} \bigg). \label{eq:MG_bound_chi_ms}
\end{equation}
In this expression, $\chi^e_\pt$ and $m^s_\pt$ are the particles'
\emph{energetic susceptibility} and \emph{saturation magnetization}, respectively. To summarize, the energy in Eq.~\eqref{eq:magnetic_energy_FB} involves a total of five magnetic parameters of the particles summarized in Table~\ref{tab:mag_params}.

\begin{table}[h!]
    \centering
    \caption{Magnetic properties of the particles.}
    \label{tab:mag_params}
    \small 
    \begin{tabular}{ p{0.1\linewidth} | p{0.13\linewidth} | p{0.13\linewidth} | p{0.15\linewidth}| p{0.1\linewidth}}
        volume fraction & energetic \,\,\,susceptibility & remanent \,\,\, susceptibiliy & magnetization saturation & coercivity
        \\[2ex]
        \hline
        $c$ \,\, [-]& $\chi^{e}_\pt$ \,\, [-] & $\chi^{r}_\pt$ \,\, [-] & $m^{s}_\pt$ \,\, [MA/m]& $b^{c}_\pt$ \,\, [T]  
    \end{tabular}
\end{table}

The saturation-type magnetization behavior of the $h$-MRE is captured by the nonlinear function $f_\pt(x)$ in Eq.~\eqref{eq:magnetic_energy_FB}, whose choice needs to be made depending on the specific saturation response of the (hard/soft) magnetic particles. Representative examples of possible choices for the functional form of $f_\pt(x)$ to describe different hard-magnetic particles have been provided in \cite{mukherjee2022} and \cite{Danas2024} and are not repeated here for brevity. In what follows, we use the inverse hypergeometric function, which reads
\begin{equation}
    f_\pt(x)=-[ \log (1 - x ) + x].
\end{equation}
\begin{figure*}[h!]
	\centering
	\includegraphics[width=0.9\textwidth]{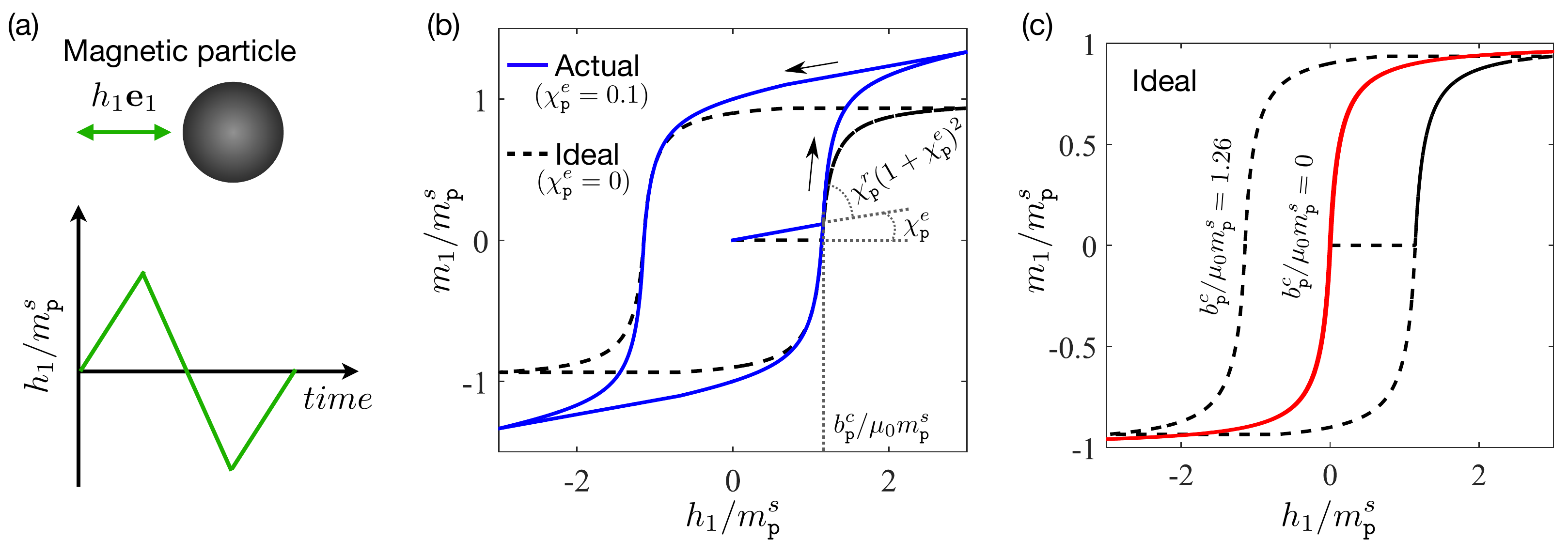}
	\caption{(a) Schematic of a magnetic particle subjected to a full cyclic magnetic load $\hvb = h_1 \mathbf{e}_1$. (b) Normalized magnetization response versus the normalized field strength. Both ideal ($\chi^e_{\mathtt{p}} = 0$; black dashed line) and actual ($\chi^e_{\mathtt{p}} = 0.1$; blue solid line) hysteresis loops are shown along with the slopes of the $m-h$ response before and after magnetic switching, \textit{i.e.}, before and after reaching the value $b^c_\pt/\muo m^s_\pt$. (c) Normalized magnetization responses for finite (black dashed line) and zero (red solid line) coercivity $b^c_\pt/\muo m^s_\pt$ lead to, respectively, hysteretic and energetic magnetization responses. (Both plots were adapted from the previous work of \cite{mukherjee2022}).}
	\label{fig:particle_hysteresis_response}
\end{figure*}

The physical interpretation of the various parameters introduced above is elucidated using Fig.~\ref{fig:particle_hysteresis_response}, which, for simplicity, refers to the case of $c=1$, corresponding to the magnetic response of pure particles (absent a matrix phase). Note that for $h$-MRE composites, which typically have $c\in(0,0.3]$ \citep{moreno2021}, the interpretation we will provide remains qualitatively unchanged. Fig.~\ref{fig:particle_hysteresis_response}a shows a schematic of a magnetic particle subjected to a full cyclic magnetic load $\hvb = h_1 \mathbf{e}_1$. Given that the particle is considered mechanically rigid, in this plot we probe the purely magnetic laws described previously. In the plot of Fig.~\ref{fig:particle_hysteresis_response}b, we observe that $\chi^e_\pt$ and $\chi^r_\pt$ control the increasing slope of the magnetization curve past the coercive field $b^c_\pt$ (to be discussed separately, below, in the context of the dissipation potential). In turn, $\chi^e_\pt$ alone controls the initial loading and unloading slope of the curve prior to reaching $b^c_\pt$ or simply the relative magnetic permeability at zero magnetic loads and a pre-magnetized state. The parameter $m^s_\pt$ defines the saturating value of the magnetization. The special case of $\chi^e_\pt=0$ corresponds to an ideal magnet, \textit{i.e.}, a magnet that reaches exactly the saturating $m^s_\pt$ value. By contrast, actual magnets continue to exhibit a slight increase of $\mvb$ up to large fields, a process that is controlled by the parameter $\chi^e_\pt$, which typically has relatively small values ranging from 0.01 to 0.2.

\paragraph{The coupling energy}
The proposed coupling free energy, $\rho_0 \Psi_{\cplt}$, in Eq.~\eqref{eq:WB_full} is written as \citep{mukherjee2022}
\begin{align}
\rho_0 \Psi_{\cplt} (I_4^{\mathtt{Hr}}, I_5^{\mathtt{BHr}}, I_5^{\mathtt{Hr}},I_6^{\mathtt{BHr}}) =
c\, \beta\, \bigg[ \mu_0 (1-2\chi^e) \Big(I_4^{\mathtt{Hr}} - I_5^{\mathtt{Hr}}\Big)  -  \dfrac{2 \chi^e}{1 + \chi^e} \Big(I_6^{\mathtt{BHr}} - I_5^{\mathtt{BHr}} \Big) \bigg]. \label{eq:FB_psi_couple_approx}
\end{align}
with the coupling parameter
\begin{equation}
\beta(c) = 19 c^2-10.4 c + 1.71, \qquad \forall c\in[0,0.3]. \label{eq:coupling_parameter_beta}
\end{equation}
The coefficients in this last equation were obtained by calibrating the analytical model to the corresponding 3D representative volume element (RVE) simulations of random distributions of magnetically hard particles in an elastomer matrix phase. This coupling parameter $\beta(c)$ may be independently calibrated against experimental data or other available numerical estimates if required. In Eq.~\eqref{eq:FB_psi_couple_approx}, we highlight the simple linear dependence of the coupling energy density on the invariants, making it possible to obtain a dual energy density in the $\Fb$-$\Hvb$ space, as discussed in detail by \cite{mukherjee2022} and not shown here for brevity. 

\paragraph{The dissipation potential}

The dissipation potential $D$ remains to be defined, which, along with the energy density $W$, will complete the constitutive relations in our model. Given that viscoelastic effects are not considered in the present manuscript (but see recent works in this direction by \cite{rambausek2021,lucarini2022,Stewart2023}), the rate-independent dissipation potential is given in terms of $\Hvbrdot$ only, such that \citep{mukherjee2019,mukherjee2021}
\begin{equation}
D(\Hvbrdot) = b^c \norm{\Hvbrdot} \ge 0,
\label{eq:Dissi_pot_def}
\end{equation}
where $\norm{.}$ denotes the standard Eulerian norm and $b^c$ is the effective \emph{coercive field} of the composite given by
\begin{equation}
b^c = b^c_\pt \bigg( \dfrac{1+\chi^e }{1 + \chi^e_\pt } \bigg)^{4/5}. \label{eq:bc_def}
\end{equation}
In this expression, $b^c_\pt$ and $\chi^e_\pt$ are the particle coercivity and energetic susceptibility, respectively, whereas $\chi^e$ has been defined in Eq.~\eqref{eq:MG_bound_chi_ms}. Typically, for a hard-magnetic composite, the effective coercivity is given by $b^c = b^c_\pt$ \citep{idiart2006}. Nonetheless, the multiplicative term of $b^c_\pt$ in Eq.~\eqref{eq:bc_def} serves as an effective correction term for an actual magnet and can be obtained from available experimental data. From the plot in Fig.~\ref{fig:particle_hysteresis_response}c, we observe that by letting $b^c_\pt \to 0$ (and by extension $b^c\to 0$), one obtains a purely energetic magnetization response (without hysteresis), relevant for the modeling of $s$-MREs. This limit, however, is not analytical, and for this reason, alternative analytical expressions were proposed in \cite{mukherjee2020} for the case of purely energetic $s$-MREs that are able to reproduce closely \citep{mukherjee2022,Danas2024} the limiting $s$-MRE response obtained by the present full dissipative model. 

The derivative of the dissipation potential in Eq.~\eqref{eq:Dissi_pot_def} with respect to $\Hvbrdot$ is non-unique at $\norm{\Hvbrdot} = 0$. Hence, we start from the Legendre-Fenchel transform of $D$, \textit{i.e.}, $D^{\ast}$, such that
\begin{equation}
D^{\ast}(\Bvbr) = \inf_{\Hvbrdot } \Big[ \Bvbr \cdot \Hvbrdot - b^c \norm{\Hvbrdot} \Big] \label{eq:def_dual_swsurf}
\end{equation}
in the rate-independent limit. In this last expression, $\Bvbr$ is the remanent $B$-like field conjugate to $\Hvbr$, such that upon the use of the condition in Eq.~\eqref{eq:def_dual_swsurf}, one can define the dissipation potential also as $D=\Bvbr \cdot \Hvbrdot\ge 0$. The minimization condition of the last expression leads to a criterion known as \emph{ferromagnetic switching surface} \citep{landis2002fully,mukherjee2019}
\begin{equation}
\Phi(\Bvbr):= \norm{\Bvbr}^2 - (b^c)^2 = 0, \label{eq:def_swsurf}
\end{equation}
which must be satisfied during the energy dissipation in a magnetic loading/unloading cycle. With Eq.~\eqref{eq:def_swsurf}, one may recast the dissipation potential $D(\Hvbrdot)$ by introducing a (non-negative) Lagrange multiplier $\dot{\Lambda}$, so that
\begin{equation}
D(\Hvbrdot) = \sup_{\Bvbr } \ \inf_{  {\dot{\Lambda} \geq 0}} \ \Big[ \Bvbr \cdot \Hvbrdot - \dot{\Lambda}\Phi(\Bvbr) \Big]. \label{eq:dissi_pot_with_lamba}
\end{equation}
In fact, substituting $\Bvbr = b^c \Hvbrdot / \norm{\Hvbrdot}$ (the minimization condition of Eq.~\eqref{eq:def_dual_swsurf}) yields exactly $D(\Hvbrdot) = b^c \norm{\Hvbrdot}$ but now with the constraint in Eq.~\eqref{eq:def_swsurf}, which must be satisfied to make the term $\dot{\Lambda}\Phi(\Bvbr)$ in Eq.~\eqref{eq:dissi_pot_with_lamba} vanish. 

The constrained dissipation potential in Eq.~\eqref{eq:dissi_pot_with_lamba} leads to the following set of equations necessary to obtain the evolution of $\Bvbr$:
\begin{equation}
\Hvbrdot = \dot{\Lambda} \dfrac{\partial \Phi}{\partial \Bvbr}, \qquad \Phi(\Bvbr) \leq 0, \qquad \dot{\Lambda} \geq 0 \qquad \text{and} \qquad \dot{\Lambda} \Phi = 0,  \label{eq:KKT_FH}
\end{equation}
which are also known as the Karush-Kuhn-Tucker (KKT) conditions \citep{Karush1939,Kuhn1951}. 

With Eq.~\eqref{eq:KKT_FH}, the evolution equations for the internal vector variable $\Hvbr$ are now fully defined, concluding our overview of the fully dissipative $h$-MRE model of \cite{mukherjee2021} and \cite{mukherjee2022}. Evidently, this model is path- and history-dependent. Its mathematical similarity to $J_2$ flow theory of plasticity allows us to resort to already well-known numerical algorithms (such as the radial return algorithm) to resolve the evolution equations and evaluate the deformation and magnetic fields under general magneto-mechanical loads. The model must be solved numerically, as is the case in all incremental elasto-plastic models in the literature. Yet, our model for MREs is explicit in the sense that all expressions are analytical and, thus, straightforward to implement in material subroutines to solve general boundary value problems \citep{rambausek2022}. Numerical implementations of the model in user element routines for Abaqus and FEniCS are available in \cite{dipayan2021uel} and \cite{rambausek2021uel}. Corresponding predictions of the proposed model may also be found in those papers and are not repeated here for conciseness. 

In closing this section, it is worth noting that it is possible to further enrich the present modeling framework to take into account more complex magnetic phenomena if deemed necessary. Possible areas for extension include switching surface shrinking during asymmetric cyclic loads, the evolution of the coercivity with temperature, or even rate or frequency effects. However, at present, there is a striking lack of experiments in the literature to study these phenomena. These experiments must be tackled first so that their results can inform future model extensions. A first effort towards this direction may be found in the earlier work of \cite{mukherjee2019}. 

\section{An energetic model for small magnetic fields after pre-magnetization}
\label{sec:An energetic model for small magnetic fields after pre-magnetization}

For small amplitudes of the applied magnetic field, where the norm of the remanent $\Bvbr$ field is smaller than the coercive value $b^c_\pt$, or simply, when $\Phi<0$ in Eq.~\eqref{eq:def_swsurf}, $\Hvbr$ remains constant (in time) throughout the actuation process even when the applied deformation is large. This observation enables us to propose a simplified purely energetic model obtained from the more general dissipation model discussed in Section~\ref{sec:The magnetic dissipation model of Mukherjee et al.}. To do so, we set a constant amplitude and direction to $\Hvbr$ 
 \citep{Moreno2022npj,Moreno2023}
\begin{equation}
\Hvbr=\boldsymbol{\mathcal{C}}, \quad \norm{\boldsymbol{\mathcal{C}}} < Q
\label{eq:Hr_const}
\end{equation}
with $Q$ denoting some real non-infinite number. Since $\Hvbr$ is constant\footnote{We precise here that $\Hvbr$ may vary in space, i.e., along the length of a beam but does not change with the applied loads in time.}, variations of the energy with respect to $\Hvbr$ are null and one may write a purely energetic model defined by the energy density
\begin{align}
W^{\mathtt{e}}(\Fb,\Bvb;\Hvbr) = \rho_0 \Psi_{\mecht}(I_1,J) + \rho_0 \Psi_{\magt}(I_5^{\mathtt{B}},I_5^{\mathtt{BHr}})+\rho_0 \Psi_{\cplt} (I_4^{\mathtt{Hr}}, I_5^{\mathtt{BHr}}, I_6^{\mathtt{BHr}}) + \dfrac{1}{2 \mu_0 J}  I_5^{\mathtt{B}}, \label{eq:WB_full_energetic}
\end{align}
where the superscript $\mathtt{e}$ stands for ``energetic''. 

In this last definition, naturally, the mechanical energy density term remains unchanged from Eq.\eqref{eq:Psi_mech}.

The purely magnetic free energy written in Eq.~\eqref{eq:magnetic_energy_FB} now simplifies to
\begin{align}
\rho_0 \Psi_{\magt}(I_5^{\mathtt{B}},I_5^{\mathtt{BHr}}) = - \dfrac{1}{2\mu_0 J} \dfrac{ \chi^{e} }{1+\chi^{e}} I_5^{\mathtt{B}} +  I_5^{\mathtt{BHr}},  \label{eq:magnetic_energy_FB_energetic}
\end{align}
with $\chi^e$ given by Eq.~\eqref{eq:MG_bound_chi_ms}. Furthermore, any terms in the energy that only involve  $\Hvbr$ (but no $\Fb$ or $\Bvb$) are constant and, thus, can be dropped. For this case of small magnetic loading, the coupled energy term is obtained by simplifying the original coupling energy density written in Eq.~\eqref{eq:FB_psi_couple_approx} to
\begin{align}
\rho_0 \Psi_{\cplt} (I_4^{\mathtt{Hr}}, I_5^{\mathtt{BHr}}, I_6^{\mathtt{BHr}}) =
c \beta(c) \bigg[ \mu_0 (1-2\chi^e) I_4^{\mathtt{Hr}}  -  \dfrac{2 \chi^e}{1 + \chi^e} \Big(I_6^{\mathtt{BHr}} - I_5^{\mathtt{BHr}} \Big) \bigg]. \label{eq:FB_psi_couple_approx_energetic}
\end{align}

\begin{remark}
Both the amplitude and direction of the internal variable $\Hvbr$ need to be prescribed in this simplified energetic model. They can be proposed heuristically based on intuitive arguments (as is usually done in the literature for simple structures). Alternatively, they need to be evaluated from an independent calculation of the magnetic dissipation to estimate the pre-magnetization profile in a magnetoelastic structure \citep{mukherjee2022}.
\end{remark}

\paragraph{Current magnetization}

Using the above expressions, one may directly evaluate the current magnetization in the $h$-MRE. Noting from the polar decomposition $\Fb=\Rb\,\Ub$ that $\Fb^{-T}=\Rb\,\Ub^{-1}$ and $\Cb^{1/2}=\Ub$, one can show using Eq.~\eqref{eq:transforms_B_and_H}$_2$ that
\begin{align}
\hvb=\Fb^{-T}\Hvb=\Fb^{-T} \dfrac{\p W^{\mathtt{e}}}{\p \Bvb}=
\dfrac{1}{\muo J}\Fb\,\Bvb- \dfrac{\chi^e}{\muo J (1+\chi^e)}\Fb \,\Bvb + \Rb \Hvbr + \dfrac{2 c \beta \chi^e}{1+\chi^e}(\Rb\,\Hvbr - \Rb\,\Cb\,\Hvbr).
\label{eq:h_Hr_b_energetic_1}
\end{align}
Next, use of Eq.~\eqref{eq:transforms_B_and_H}$_1$ leads to
\begin{align}
\hvb=\dfrac{1}{\muo}\bvb- \dfrac{\chi^e}{\muo (1+\chi^e)}\bvb + \Rb \Hvbr + c \beta \dfrac{2 \chi^e}{1+\chi^e}(\Rb\,\Hvbr - \Rb\,\Cb\,\Hvbr).
\label{eq:h_Hr_b_energetic_2}
\end{align}
A direct comparison between this last relation and the definition for the current magnetization in Eq.~\eqref{eq:magnetization_definition} yields 
\begin{align}
\mvb=\dfrac{\chi^e}{\muo (1+\chi^e)}\bvb - \Rb \Hvbr - \dfrac{2 c\, \beta \,\chi^e}{1+\chi^e}(\Rb\,\Hvbr - \Rb\,\Cb\,\Hvbr),
\label{eq:Hr-m_fully_coupled model}
\end{align}
which provides a relation between $\mvb$, $\bvb$, and $\Hvbr$ in the energetic, coupled model proposed here. 

Certain observations follow:
\begin{itemize}

\item  In the absence of an applied magnetic field $\bvb$, the current magnetization $\mvb$ is only a function of $\Hvbr$ and the rotation part of the deformation gradient $\Rb$. 

\item In the case of non-ideal $h$-MRE, where $\chi^e \neq 0$ (see the example in Fig.~\ref{fig:particle_hysteresis_response} or \cite{Stepanov2017} and \cite{mukherjee2019} for relevant experimental data), $\mvb$ evolves with the (applied) Eulerian magnetic flux $\bvb$ (as shown in Fig.~\ref{fig:particle_hysteresis_response}b) with a slope that is not $\muo$ but depends on $\chi^e$. In turn, $\chi^e$ is a function of the particle permeability and volume fraction as defined in Eq.~\eqref{eq:MG_bound_chi_ms}. Note, however, that in realistic $h$-MREs, $\chi^e$ is rather small (\textit{i.e.}, $\chi^e \sim 0.01-0.2$) and, thus, may be neglected in several cases of interest \citep{zhao2019}.

\item We observe that $\mvb$ is not a proper internal variable since, still in the general case of non-ideal $h$-MREs, $\mvb$ depends also on $\bvb$ and not only on $\Hvbr$. 

\item The last term in Eq.~\eqref{eq:Hr-m_fully_coupled model} implies that for non-ideal magnets with $\chi^e \neq 0$, $\mvb$ will change with the application of a stretch in the full energetic model defined in Eq.~\eqref{eq:WB_full_energetic}. Nevertheless, this last term is substantially smaller than the first two, since in actual $h$-MREs, $\chi^e \sim 0.01-0.2$ and $c \beta$ are of a similar order when $c\leq 0.3$. This implies that this term is, in practice, very small and will essentially lead to negligible changes in magnetization amplitude for stretches as large as three (\textit{i.e.}, 200\% strain). As such, we reiterate the observation that $\mvb$ is a function of $\Hvbr$ only and \emph{nearly} stretch independent in most cases of practical interest. This term, however, derives from the coupling part of the energy, which is essential for the prediction of magnetostriction at large magnetic fields.
\end{itemize}

\subsection{Special case of uncoupled, ideal \textit{h}-MRE}
\label{subsec:Special case of uncoupled, ideal $h$-MRE}

The case of an ideal magnetic (energetic) $h$-MRE, i.e., one with ideal magnetic particles (see Fig.~\ref{fig:particle_hysteresis_response}b) has been extensively used in the literature owing mainly to the seminal work of \cite{zhao2019}, which focuses, however, on slender MRE structures subjected mainly to bending deformations. Next, we provide a detailed discussion of how the full energetic model in Eq.~\eqref{eq:WB_full_energetic} can be simplified for the case of ideal magnetic $h$-MREs and comment on the implications of this simplification.

We start by using an uncoupled magneto-mechanical energy, an approximation that was considered in the earlier models of \cite{zhao2019} and \cite{Yan2023}, as well as more recently in \cite{Moreno2023}, and can be implemented by setting $\beta=0$ in Eq.~\eqref{eq:FB_psi_couple_approx_energetic}, leading to
\begin{align}
    \rho_0 \Psi_{\cplt}=0.
\end{align}
Here, the notion of an uncoupled response is meant in the sense of the homogenization analysis carried out by \cite{danas2017effective}, where it was shown that under the application of a Eulerian magnetic field $\bvb^\mathtt{app}$, the purely magnetic energy in Eq.~\eqref{eq:magnetic_energy_FB_energetic} induces no net intrinsic magnetostriction of the MRE. Nonetheless, by \textit{uncoupled}, one should not be confused with the apparent magneto-mechanical coupling induced between a magnetic structure (even with negligible magnetostriction) and the surrounding magnetic field; this is, in fact, a strong effect with the rotation of a metallic needle in a compass the canonical example. While the compass needle has negligible magnetostrictive strains, the structure interacts with the surrounding magnetic field to indicate the direction of the Earth's magnetic poles. 

Another approximation required to recover the magnetic response of an ideal $h$-MRE consists in setting $\chi^e_\pt=0$, and thus $\chi^e=0$, in the energetic model of Eq.~\eqref{eq:WB_full_energetic}. This approximation simplifies the magnetic energy in Eq.~\eqref{eq:magnetic_energy_FB_energetic} to 
\begin{align}
\rho_0 \Psi_{\magt}(I_5^{\mathtt{BHr}}) =I_5^{\mathtt{BHr}}=\Bvb \cdot \Cb^{1/2}\Hvbr.  \label{eq:Psi_mag_FB_energetic_ideal}
\end{align}
The effect of setting $\chi^e_\pt=0$ has been discussed in Section~\ref{subsec:Energy densities and dissipation potential} (cf. Fig.~\ref{fig:particle_hysteresis_response}b) and serves to impose that the relative magnetic permeability of a pre-magnetized $h$-MRE is exactly equal to unity in the absence of an applied magnetic field. 

Gathering the approximations and statements provided above, we obtain the following simplified energetic model for an ideal $h$-MRE:
\begin{align}
W^{\mathtt{e,R}}(\Fb,\Bvb;\Hvbr) &= \rho_0 \Psi_{\mecht}(I_1,J) + I_5^{\mathtt{BHr}} + \dfrac{1}{2 \mu_0 J}  I_5^{\mathtt{B}}\nonumber \\[1ex]
	&= \rho_0 \Psi_{\mecht}(I_1,J) + \Bvb \cdot \Cb^{1/2}\Hvbr + \dfrac{1}{2 \mu_0 J} \Bvb\cdot \Cb \,\Bvb.
\end{align}
Following the same steps presented in Eq.~\eqref{eq:h_Hr_b_energetic_1}, or simply setting $\chi^e=0$ in Eq.~\eqref{eq:Hr-m_fully_coupled model}, one can readily show that in the present case of an uncoupled, ideal $h$-MRE, the current magnetization is 
\begin{align}
\mvb= - \Rb \Hvbr.
\label{eq:Hr-m_ideal model}
\end{align}
Importantly, given the starting assumptions, especially in light of Eq.~\eqref{eq:Hr_const}, we observe that the magnitude of the current magnetization, $|\mvb|=|\Hvbr|$, remains unchanged with the application of a small applied (external) magnetic field and is entirely stretch-independent. Even in this simplified case, $\mvb$ cannot formally serve as an internal variable since it depends on the rotation $\Rb(\Fb)$, which itself is a function of the deformation gradient. By contrast, it is evident that $\mvb$ will exhibit properties similar to $\Hvbr$, which, by definition, is an internal variable.

For the purpose of comparing this uncoupled model with the models of \cite{Yan2023} (cf. Section~\ref{sec:The RM energetic model of Yan et al.}) and \cite{zhao2019} (cf. Section~\ref{sec:Connections and differences between the models}), it is helpful to define an energy per unit current volume $w^{\mathtt{e,R}}=W^{\mathtt{e,R}}/J$, which, together with the result Eq.~\eqref{eq:Hr-m_ideal model}, reads
\begin{align}
W^{\mathtt{e,R}}(\Fb,\bvb,\Hvbr) &= \rho_0 \Psi_{\mecht}(I_1,J)+J\,\Rb\Hvbr\cdot \bvb + \dfrac{J}{2 \mu_0} \bvb\cdot\bvb,
\label{eq:WeR_Hrb_definition}
\end{align}
or equivalently
\begin{align}
w^{\mathtt{e,R}}(\Fb,\bvb,\Hvbr) &= \dfrac{\rho_0}{J} \Psi_{\mecht}(I_1,J)-\mvb\cdot \bvb + \dfrac{1}{2 \mu_0} \bvb\cdot\bvb.
\label{eq:weR_mb_definition}
\end{align}
It is straightforward to show then that 
\begin{equation}
\label{eq:dwder_h-b}
\dfrac{\p w^{\mathtt{e,R}}}{\p \bvb}= -\mvb+ \dfrac{1}{\muo} \bvb = \hvb.
\end{equation}

\begin{remark}
The derivatives of both the coupled, $W^\mathtt{e}$, and the uncoupled $W^{\mathtt{e,R}}$ energies with respect to $\Fb$ lead to a symmetric total Cauchy stress, which may be split to a sum of an energetic mechanical stress and a Maxwell stress. Those expressions have been discussed extensively in \cite{mukherjee2022} and are not reported here for brevity.
\end{remark}

\section{The energetic model of \cite{Yan2023}} 
\label{sec:The RM energetic model of Yan et al.} 

Following an approach similar to that of \cite{zhao2019}, \cite{Yan2023} proposed a magneto-elastic model for $h$-MREs valid for small magnetic fields around a pre-magnetized state. Inspired by their own experimental results, the model in \cite{Yan2023} does, however, take into consideration the stretch-independency of $\mvb$, contrary to that of \cite{zhao2019}, which does not. 

To facilitate the comparison between these existing models and the framework presented in this note, we have updated their notation in the summary of the model of \cite{Yan2023} provided next. The behavior of a bulk, magnetically ideal $h$-MRE under a small applied magnetic field and a fully pre-magnetized state is described by a Helmholtz free energy density, summing an elastic mechanical part, $\rho_0 \Psi_\mecht$ and a magnetic part, $\rho_0 \Psi^{\mathtt{RM}}_\magt$,\footnote{In the original studies of \cite{Yan2023} and \cite{Yan2021ijss}, the notation $U^\texttt{e}\equiv\rho_0 \Psi_\mecht$ and $U^\texttt{m}\equiv\rho_0 \Psi^{\mathtt{RM}}_\magt$ was used.}, such that
\begin{align}
w^{\mathtt{RM}}(\Cb,\bvb,\mvb) = \dfrac{\rho_0}{J} \Psi_{\mecht}(I_1,J) + \dfrac{\rho_0}{J} \Psi^{\mathtt{RM}}_{\magt}(\mvb,\bvb). \label{eq:WRM_def}
\end{align}
For the mechanical part, the authors used a simple quasi-incompressible neo-Hookean model of the form 
\begin{equation}
	\dfrac{\rho_0}{J} \Psi^{\mathtt{RM}}_\mecht(I_1)=\dfrac{G}{2}(J^{-2/3}\Fb^T\cdot \Fb-3) + \dfrac{K}{2}(J-1)^2.
	\label{eq:Psi_mech_RM}
\end{equation}
In this expression, the shear, $G$, and bulk, $K$, moduli of the $h$-MRE may be either measured directly from experiments or estimated using the homogenization result in Eq.~\eqref{eq:Psi_mech}. For the experimentally relevant case of a neo-Hookean matrix phase and mechanically rigid particles, use of the homogenization estimates leads to \citep{lopezpamies2013,Luo2023}
\begin{equation}
	G= \dfrac{G_\mt}{(1-c)^{\frac{5}{2}}}, \quad \text{and} \quad K=\dfrac{K_\mt}{2 (1-c)^6},
\end{equation}
where $G_\mt$ is the shear modulus of the polymer matrix and $c$ is the volume fraction of the rigid NdFeB particles. 

The corresponding magnetic energy reads 
\begin{align}
\dfrac{\rho_0}{J} \Psi^{\mathtt{RM}}_{\magt}(\mvb,\bvb)=- \mvb \cdot \bvb.
\label{eq:Psi_mag_RM}
\end{align}
Subsequently, the authors, making use of direct experimental evidence and numerical results of \cite{mukherjee2021}, described the pre-magnetized state using a magnetization measure $\Mvb$, which is \emph{constitutively} linked to the current magnetization $\mvb$ via
\begin{equation}
	\mvb = J^{-1} \Rb \Mvb, \qquad \Mvb \in \Rthr.
	\label{eq:m-RM_relation}
\end{equation}
The vector $\Mvb$ serves to describe the pre-magnetization state, which remains fixed during actuation. More importantly, the fact that $\Mvb$ is related to the current $\mvb$ via $\Rb$ (and not via $\Fb$) implies that it lies at an intermediate stretch-free configuration. By direct comparison with the energetic models presented in Section~\ref{subsec:Special case of uncoupled, ideal $h$-MRE}, one can connect the internal variable $\Hvbr$ and $\Mvb$ simply by setting
\begin{equation}
	\Hvbr=-J^{-1} \Mvb.
	\label{eq:Hr-M_relation}
\end{equation}
As such, the model of \cite{Yan2023} turns out be exactly equivalent (up to the background term $1/\muo \bvb\cdot\bvb$) to the simplified, energetic version of the \cite{mukherjee2021} and \cite{mukherjee2022} model presented in Eq.~\eqref{eq:weR_mb_definition}.

\section{The \cite{zhao2019} model versus the energetic models of \cite{mukherjee2021} and  \cite{Yan2023} }
\label{sec:Connections and differences between the models}

In the original work of \cite{zhao2019}, the authors considered a remanent magnetic field\footnote{The remanent magnetic field was denoted as $\Bvb^r$ in their article.}, $\bvb^r$, which has the characteristics of an internal variable similar to $\Hvbr$ defined in Eq.~\eqref{eq:Hvbr_def}. They also assumed that $\bvb$ and $\hvb$ remain linear and with a slope $\muo$ at small applied fields (\textit{i.e.}, ideal $h$-MREs), which is a fair assumption since the relative permeability in such materials is close to unity, as already discussed in Section~\ref{sec:Introduction and problem definition}. Together with the very definition of the current magnetization in Eq.~\eqref{eq:magnetization_definition}, this assumption yields
\begin{equation}
	\hvb=\dfrac{1}{\muo}(\bvb-\bvb^r) \qquad \Rightarrow \qquad \bvb^r=\muo \mvb.
 \label{eq:br_equal_m}
\end{equation}
Subsequently the field $\bvb^r$ is pulled back to obtain a Lagrangian remanent field, $\Bvb^r$ (denoted as $\tilde{\Bvb}^r$ in their article), defined as 
\begin{align}
    \Bvb^r= J \Fb^{-1} \bvb^r=J \muo \Fb^{-1} \mvb,
\end{align}
which is directly related to the remanent field in the $h$-MRE after strong pre-magnetization. By contrast, during pre-magnetization the above relation implies
\begin{align}
    \dfrac{1}{J} \Fb\Bvb^r= \bvb^r=\muo \Mvb,
    \label{eq:br_eq_M_premag}
\end{align}
and leads to $\hvb=\mathbf{0}$ during the pre-magnetization operation.

Subsequently, $\Bvb^r$ is chosen as the main independent state variable in the problem, leading to the following definition of the energy density:
\begin{align}
w^{\mathtt{FM}}(\Cb,\bvb,\mvb) = \dfrac{\rho_0}{J} \Psi_{\mecht}(I_1,J) - \dfrac{1}{J\mu_0} \Fb \Bvb^r\cdot \bvb. \label{eq:WFM_def}
\end{align}
Importantly, this form of the energy density implies that the constant Lagrangian remanent field\footnote{The Lagrangian remanent field, $\Bvb^r$, may be identified with the pre-magnetized state by setting $\Bvb^r=\muo \Mvb$ in the notation of \cite{Yan2023}.}, $\Bvb^r$, directly leads to strong stretch-dependent $\bvb^r$ and $\mvb$. As a consequence, it follows that for arbitrary stretches $\Ub$ (from the polar decomposition $\Ub=\Rb^T\,\Fb$), their model predicts that the current magnetization could become substantially larger than the saturation magnetization of the $h$-MRE since it evolves with the deformation gradient $\Fb$.

By contrast, \cite{zhao2019} have only implemented and used the proposed $\Fb$-$\Bvb^r\equiv \Fb$-$\Mvb$ model in situations of very small stretches, such as bending of an actuated beam or plate, which are both slender geometrically and thus use of the inextensibility condition results naturally to a stretch independent magnetization response. If, however, the plate, shell or beam is stretched, as is the case in the experimental results of \cite{Yan2023} and the recent studies of \cite{Stewart2023} and \cite{Zhang2023}, then significant discrepancies may be observed between the $\Fb\Mvb$ model of \cite{zhao2019} and the $\Rb\Mvb$ model of \cite{Yan2023} and that of \cite{mukherjee2021}.

\begin{figure*}[h!]
	\centering
	\includegraphics[width=0.9\textwidth]{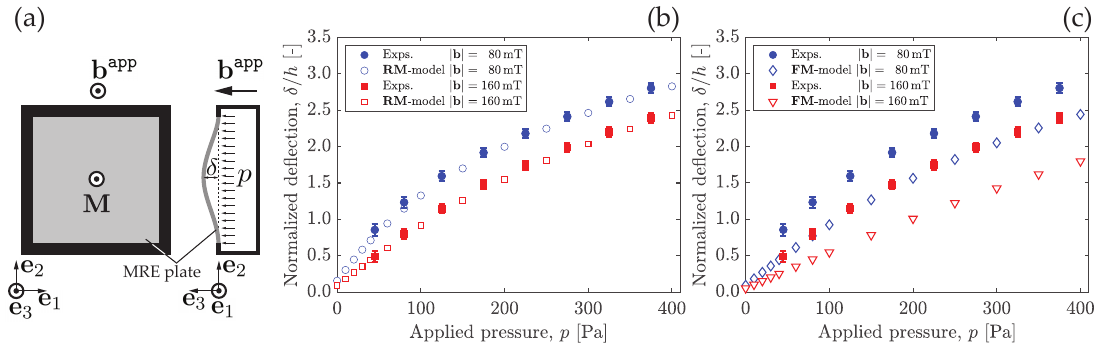}
	\caption{Results reproduced from \cite{Yan2023} for experiments and FEM simulations, the latter for both the $\textbf{RM}$ and the $\textbf{FM}$ models, for a fully clamped $h$-MRE plate under combined pressure and magnetic loading. (a) Schematic diagram with a top view (into the $\mathbf{e}_1$-$\mathbf{e}_2$ plane) and side view (into the $\mathbf{e}_2$-$\mathbf{e}_3$ plane) of the system. (b,c) Normalized deflection, $\delta/h$ versus the applied pressure $p$, for two values of the amplitude of the applied magnetic field, $|\mathbf{b}|=\{80,\, 160\}\,$mT. The experimental results (solid symbols) are common to both plots. (b) Comparison between the experimental results and 3D FEM-simulations (open symbols) using the $\mathbf{RM}$ model: circles and squares for 80\,mT and  160\,mT, respectively. (c) Comparison between the experimental results and  3D FEM-simulations (open symbols) using the $\mathbf{FM}$ model: open diamonds and triangles for 80\,mT and and 160\,mT, respectively. (This figure was adapted from the previous work of \cite{Yan2023}).}
\label{fig:yan_et_al_magneticplates}
\end{figure*}

We proceed by reviewing the subset of results from \cite{Yan2023} that are relevant to, and provide evidence for, the discussion in the paragraph above. For convenience, these results are reproduced in Fig.~\ref{fig:yan_et_al_magneticplates}. As shown in the schematic diagram of Fig.~\ref{fig:yan_et_al_magneticplates}a, their system comprised a square-shaped, thin, $h$-MRE plate (side length $L=25\,$mm and thickness $h=\,505\mu$m) that was clamped on all of its 4 sides. The plate was magnetized along $\mathbf{e}_3=\Mvb/\norm{\Mvb}$, which is also the direction of the applied magnetic field $\mathbf{b}^{\texttt{app}}$. A pneumatic chamber induces an internal pressure $p$. The deflection of the plate along $\mathbf{e}_3$ is quantified by the maximum value, $\delta$, at its center. We point the reader to the original manuscript of \cite{Yan2023} for additional details (\textit{e.g.}, fabrication, additional physical properties, and experimental protocol). In Fig.~\ref{fig:yan_et_al_magneticplates}b,c, we reproduce their experimental data (closed symbols) for the measured normalized deflection, $\delta/h$ as a function of the applied pressure, $p$, for two values of the amplitude of the applied magnetic field, $|\mathbf{b}|=\{80,\, 160\}\,$mT. At the same value of the applied pressure $p$, a higher magnetic field resists the plate deflection. The corresponding 3D FEM results (open symbols) are presented in Fig.~\ref{fig:yan_et_al_magneticplates}b using the $\mathbf{RM}$ model and in Fig.~\ref{fig:yan_et_al_magneticplates}c using the $\mathbf{FM}$ model. The authors found that the $\mathbf{RM}$ model provides predictions in excellent agreement with the experimental data, whereas the $\mathbf{FM}$ model does not. The pneumatic loading induces non-negligible stretching deformation of the plate's mid-surface, calling for an appropriate description of the magnetization of the deformed plate according to Eq.~\eqref{eq:m-RM_relation} ($\mathbf{RM}$ model) and that the $\mathbf{FM}$ model by \cite{zhao2019} is inappropriate in this case. Note that \cite{Yan2023}, as well as other studies mentioned in Section~\ref{sec:Introduction and problem definition}, also investigated several other cases of slender structures that effectively behaved as inextensible, for which the predictions from the $\mathbf{FM}$ model were satisfactory.

\section{Discussion and limitations of the simplified models}

We proceed by commenting on the fact that the models of \cite{zhao2019}, \cite{Yan2023} and many others thereafter in the literature neglect the background energy term $\Bvb\cdot \Cb\,\Bvb/2 \muo J$; cf. the last term in Eq.~\eqref{eq:weR_mb_definition}. One may directly add this term in Eqs.~\eqref{eq:WRM_def} and~\eqref{eq:WFM_def}, similar to the energy definitions in Eqs.~\eqref{eq:WB_full_energetic},~\eqref{eq:weR_mb_definition}, and \eqref{eq:WeR_Hrb_definition}. However, in that case, $\Bvb$ (or $\bvb$) becomes an unknown field with prescribed boundary conditions and needs to be solved for, together with the displacement field. Instead, dropping this background magnetic term may be justifiable, but \textit{only} under specific conditions, which are summarized as follows\footnote{See also the related discussion in \cite{Sharma2020}.}:
\begin{enumerate}[i.]
\item the externally applied magnetic field is spatially uniform (\textit{i.e.}, it is applied far from the specimen),

\item the MRE geometry is such that the $\Bvb$ and $\Hvb$ fields are (or may be approximated to be) uniform in the specimen,  

\item the Eulerian magnetic flux $\bvb$ or magnetic strength $\hvb$ in the specimen can be assumed to be equal to the externally applied Eulerian magnetic flux $\bvb^\texttt{app}$ or $\hvb^\texttt{app}$, respectively,

\item the Eulerian magnetic strength $\hvb$ or magnetic flux $\bvb$ during pre-magnetization are approximately zero (for applied $\bvb^\texttt{app}$ or $\hvb^\texttt{app}$, respectively), such that relation in Eq.~\eqref{eq:br_eq_M_premag} holds.

\end{enumerate}

In the specific situations stated above, due to the normal continuity of $\Bvb$ and the tangential continuity of $\Hvb$, it can be readily shown that the leading order tractions due to Maxwell stresses resulting from the background magnetic energy term, $\Bvb\cdot \Cb\,\Bvb/2 \muo J$, inside the MRE and the surrounding air are identically equilibrated. As such, this term has no net impact on the \emph{mechanical} response of the MRE (soft or hard), and the torque exerted upon the structure via the magnetic field is $\bvb^\texttt{app} \times (\muo \Mvb)$. Note that, in actual experimentally relevant situations, this assumption may introduce errors, which can be negligible or substantial, depending on the geometry of the device and/or the MRE. Certain important counter-examples of practical relevance are discussed next.

Near corners of the MRE structure, the magnetic fields cannot be uniform by standard physical arguments (\textit{e.g.}, the requirement of a divergence-free $\Bvb$ and a curl-free $\Hvb$), also known as Fringe effects. In bulk specimens, it was shown for small strains by \cite{brown1966}, and for finite strains by \cite{Lefevre2017}, that the mechanical and magnetic fields are not uniform inside an MRE specimen, even when the far-applied pre-magnetization or actuation magnetic field is uniform. This non-uniformity becomes less pronounced in slender structures except near the edges of the specimen.

Another case where assumptions (iii) and (iv) are not satisfied occurs when a uniformly pre-magnetized (along the long axis) $h$-MRE beam is at an angle with respect to the applied actuation magnetic field. In this configuration, the magnetic flux in the solid cannot be the same as that of the external air since a non-zero component exists along the beam's long axis. At one hand, this component will affect the shear strains and stresses in the beam. On the other hand, this component is not expected to affect strongly the overall deflection of the beam at the initial stages of deformation if the main deformation mode is pure bending. Yet, this non-zero component of the resulting magnetic flux during actuation along the beam's long axis may strongly affect the response of the system at large applied fields and more complex structures or more complex pre-magnetization profiles \citep{mukherjee2022}.

Another example where the background magnetic energy cannot be neglected can be found in the work of \cite{psarra2019} studying a system comprising a thin $s$-MRE film attached to a passive elastic substrate. Their experimental and numerical investigation clearly showed that the shear stresses in the film lead to a crinkling instability pattern with pronounced angular out-of-plane protrusions similar to the ferrofluid Rosensweig instability \citep{rosensweig1967}. In that study, the proper resolution of the surrounding magnetic energy and the interaction between neighboring wrinkles/crinkles turned out to be of critical importance to correctly capture the experimentally observed deformation pattern.

Similarly, condition (iv) in the above list can only be valid when the pre-magnetization field has a simple uniform distribution along the principal directions of the slender structure. When, for instance, the pre-magnetization is conducted on pre-curved slender structures \citep{ren2019}, despite being slender, both $\bvb$ and $\hvb$ fields are non-null and thus neither of them can be directly and analytically identified with the external applied magnetic fields. Furthermore, in those cases, the pre-magnetized state does not correlate directly with the pre-deformed shape of the structure \citep{mukherjee2022}, and thus, a full simulation of the boundary value problem (BVP) is required to estimate the actual pre-magnetization state. 

In actual magnetic setups, the applied pre-magnetization or actuating magnetic fields may be non-uniform \citep{dorn2021}. A rather critical assessment of several popular setups and their influence on the MRE response has been recently carried out in \cite{Moreno2023}. In such cases, none of the above conditions holds true, calling for simulations of the full BVP. \cite{Yan2021ijss}  and \cite{Sano2022a} have also addressed the general case of non-uniform fields with a constant gradient (albeit using the previous $\mathbf{FM}$ model only valid for negligible stretching) by incorporating the magnetic body force induced by the field gradient and implementing it in FE packages. \cite{Yan2021ijss} focused on the planar (2D) deformation of geometrically nonlinear hard-magnetic beams, whereas \cite{Sano2022a} studied the 3D deformation of hard-magnetic rods following a Kirchhoff rod theory framework. Given the slenderness of these beams and rods, the specific loading conditions, and the relatively small magnitude of the applied actuating magnetic fields, both of these systems remained effectively inextensible. As such, it was appropriate for the authors to employ the $\mathbf{FM}$ model, in lieu of the $\mathbf{RM}$ one, and also without a need to consider the $\Psi_{\cplt}$ term in the energy density of Eq.~\eqref{eq:WB_full_energetic}.

For more complex and multi-component $h$-MRE structures, which may exhibit self-interactions, one should take into account the surrounding air in the modeling approach. The chosen approach may vary depending on the BVP at hand, the complexity of the geometry, and the convergence properties of the numerical scheme. We point the interested reader to a number of recent efforts along this direction: the staggered approach \citep{pelteret2016} (see also \cite{rambausek2022} for improvements), the penalty method of constraining the air-solid boundary node sets \citep{psarra2017,psarra2019}, and the more straightforward use of a very soft mechanical law for the air (see for instance \cite{dorn2021} and improvements in \cite{rambausek2022} and \cite{moreno2022}). More recently, a non-local approach coupling interacting faces in a beam structure by use of dipole-dipole interactions has been proposed by \cite{Sano2022b} to rationalize previously unexplained experimental observations in \cite{Sano2022a}. Finally, two additional promising approaches, not discussed here, have been proposed in \cite{Rambausek2023}, where a proper treatment of the Maxwell stress at the interface between the magnetoelastic solid and the air allows to eliminate the spurious modes present in such problems and allow for good convergence. 

Finally, and from a more mathematical point of view, we note that the omission of the background magnetic energy may lead to non-symmetric stress measures and incomplete descriptions of the magnetic constitutive relations. For instance, we observe that neglecting the term $\bvb\cdot\bvb/2 \muo$ in Eq.\eqref{eq:weR_mb_definition} would lead to an incomplete definition for the magnetic field strength in Eq.\eqref{eq:dwder_h-b}; \textit{i.e.}, $\partial w^{\mathtt{e,R}}/\p \bvb= -\mvb$ and not $\hvb$.

\section{Conclusion}
\label{sec:conclusion}

In this study, we aimed to elucidate several crucial aspects concerning the modeling of $h$-MREs (but also relevant for $s$-MREs) and, in particular, the stretch-independence of their magnetization response, which has been observed experimentally. Such a property needs to be taken into account appropriately in the modeling so as to avoid inaccurate or nonphysical predictions (\textit{e.g.}, a magnetization that is larger than the magnetization saturation of the MRE) that may result when the tested MRE is subjected to pre-stretching or pre-stressing. We have shown that the fully dissipative model of \cite{mukherjee2021} may be reduced, under certain physically sound assumptions, to the energetic model of \cite{Yan2023}, but not that of \cite{zhao2019}. The former two models were shown to be in agreement with experiments on slender structures. Furthermore, the more complete dissipative model of \cite{mukherjee2021} was also shown to be in very good agreement with full-field numerical simulations of representative volume elements of $h$-MREs subjected to finite stretching and magnetic fields and not only pure bending. 

As an additional outcome of the present analysis, we have also shown that the magnetization has the properties of an internal variable, and as such, it is subject to constitutive assumptions. Specifically, the use of a direct pull-back or push-forward of the current magnetization may lead to incorrect predictions that are inconsistent with experimental measurements or full-field simulations. The reason lies in the fact that in a finite strain formulation, the magnetization is constitutively defined in the current configuration \citep{kankanala2004finitely}, while it has no particular physical meaning in the undeformed configuration. Any attempt to use such a Lagrangian measure for the magnetization as an independent variable can lead to nonphysical measures of the latter in the current configuration. Such is the case for the $\Fb\Mvb$ model, originating by such a pull-back operation. 

Finally, neglecting the background magnetic energy, an assumption typically made for simplicity of the analysis, may only be made in specific cases involving slender structures that are pre-magnetized uniformly and subjected to uniformly applied magnetic fields or with specific gradients. Even then, the reduced models may be able to predict accurately the mechanical response of the material or structure but, by construction, are incomplete and thus incapable of describing the resulting magnetic response and its evolution during the loading process. 

We close by noting that the observed stretch-independence of the magnetization in $h$-MREs is also a feature in $s$-MREs; this was shown unambiguously in the earlier work of \cite{danas12} and a considerable effort to take it into account was made in their modeling approach, as well as in subsequent models \citep{mukherjee2020}. We hope that the present study will yield a better understanding of and further the design of programmable materials and soft robots comprising components made of $h$-, $s$-, or hybrid MRE materials.


\section*{Acknowledgements}
The authors would like to thank Prof. Dipayan Mukherjee for his careful reading of the manuscript and his valuable suggestions. K.D. would like to acknowledge support from the European Research Council (ERC) under the European Union’s Horizon 2020 research and innovation program (grant agreements 636903 and 101081821).

\bibliographystyle{abbrvnat}
\bibliography{myref}
\end{document}